\documentclass[a4paper,10pt,dvipsnames]{IEEEtran}
\usepackage{graphics, graphicx}
\usepackage{amssymb,amsmath}
\usepackage{amsmath}
\usepackage[rm]{subfigure}   
\usepackage{threeparttable}  
\usepackage{color}
\usepackage{url}
\usepackage[T1]{fontenc}
\usepackage{algorithm}
\usepackage{cite}
\usepackage{multirow}
\usepackage[normalem]{ulem}
\usepackage{float}
\usepackage{flafter}
\usepackage{algorithmic}
\usepackage{mathtools}

\usepackage{xspace,colortbl}
\usepackage{tikz}

\usepackage{amsmath}

\usepackage{indentfirst}
\usepackage{setspace}

\usepackage{xcolor}
\usepackage{tikz}
\usetikzlibrary{shapes,arrows,backgrounds,fit}
\usepackage{pgfplots}
\tikzstyle{surround} = [thick,draw=black,rounded corners=2mm]
\tikzset{%
	block/.style    = {draw, rectangle, rounded corners, fill=none, minimum size=2em, minimum width=10em},
	decision/.style = {draw, diamond, rounded corners, fill=none, aspect=2},
	circulo/.style  = {draw, circle, node distance = 1.5cm}
}

\usepackage{tikz}
\definecolor{gold}{rgb}{0.85,.66,0}

\newcommand{\jcm}{\textcolor{black}}
\newcommand{\ta}{\textcolor{black}}

\begin{document}

\title{Exploring the Non-Overlapping Visibility Regions in XL-MIMO Random Access Protocol}
\author{\jcm{José Carlos Marinello Filho}, 
{Glauber Brante}, {Richard Demo Souza}, {Taufik Abrão}\\
\vspace{-3mm}
\thanks{This work was supported in part by the National Council for Scientific and Technological Development (CNPq) of Brazil under Grant  310681/2019-7.}
\thanks{J. C. Marinello is with the Electrical Engineering Department, Federal University of Technology PR, Cornélio Procópio, PR, Brazil. {\scriptsize jcmarinello@utfpr.edu.br}} 
\thanks{G. Brante is with the Department of Electrotechnics, Federal University of Technology PR, Curitiba, PR, Brazil. {\scriptsize gbrante@utfpr.edu.br}}
\thanks{ R. D. Souza is with the Electrical and Electronics Engineering Department, Federal University of Santa Catarina (UFSC), Florianópolis, SC, Brazil. {\scriptsize richard.demo@ufsc.br}} 
\thanks{T. Abrão is with Electrical Engineering Department, State University of Londrina, PR, Brazil. {\scriptsize taufik@uel.br}}
}

\maketitle

\begin{abstract}
The {recent} extra-large scale massive multiple-input multiple-output (XL-MIMO) systems are seen as a promising technology for providing very high data rates in increased user-density scenarios. Spatial non-stationarities and visibility regions (VRs) appear across the XL-MIMO array, since its large dimension is of the same order of the distances to the user-equipments (UEs). Due to the increased density of UEs in typical applications of XL-MIMO systems and the scarcity of pilots, the design of random access (RA) protocols and scheduling algorithms become challenging. In this paper, we propose a joint RA and scheduling protocol, {namely} non-overlapping VR XL-MIMO (NOVR-XL) RA protocol, which takes advantage of the different VRs of the UEs for improving RA performance, besides of seeking UEs with non-overlapping VRs to be scheduled in the same payload data pilot resource. Our results {reveal} that the proposed scheme achieves significant gains in terms of sum-rate compared with traditional {RA} schemes, as well as reducing access latency and improving connectivity performance as a whole.
\end{abstract}
\begin{IEEEkeywords}
Random access protocol, grant-based protocols, massive MIMO, XL-MIMO, non-stationarity, visibility region (VR), scheduling.
\end{IEEEkeywords}

\section{Introduction}\label{sec:intro}
Unusual channel conditions are seen in the recent extra-large scale massive multiple-input multiple-output (XL-MIMO) technology \cite{Tufvesson20a}. By distributing the antenna elements in a wide space, like in the facades of buildings, stadiums, or shopping malls, the distance between user equipments (UEs) and the base station (BS) antennas might become of the same order of the array dimension, giving rise to spatial non-stationarities, in which the received signal energy of a given UE varies significantly accross the array \cite{Tufvesson20a, Wang20, Han_2020, Ali_2020, Souza_2021, Croisfelt_2021}. Besides, since different regions of the large-aperture array see different sets of scatterers, obstacles, and UEs, the concept of visibility regions (VRs) is introduced in this channel model \cite{Ali_2020}. Through the concept of VRs, some antennas, sometimes grouped in subarrays (SAs), are visible by a given UE, \emph{i.e.}, receive significant signal energy, while other antennas are not \cite{Tufvesson20a, Wang20, Han_2020, Ali_2020, Souza_2021, Croisfelt_2021}.

Some envisioned services within the future sixth generation of wireless networks (6G) will demand a highly efficient method to deal with massive UE access attempts, as massive Machine Type Communication (mMTC) and crowded Mobile Broadband (cMBB) \cite{Carvalho_2017}. In these scenarios, the number of available pilot sequences is lower than the number of UEs. Therefore, implementing efficient random access (RA) protocols becomes essential to allow coherent communication. An important RA protocol for spatially stationary massive MIMO systems is proposed in \cite{Bjornson_2017}, namely the strongest user collision resolution (SUCRe) protocol. This protocol aims to resolve RA pilot collisions by selecting the UE with the strongest signal gain to repeat its pilot after the collision, in a completely decentralized and uncoordinated fashion \cite{Bjornson_2017}.

\jcm{Extending \cite{Bjornson_2017}, the authors in \cite{SUCR-IPA} propose an idle pilot access (SUCR-IPA) protocol, which allows a fraction of the users that failed to access the BS to re-select their pilots by means of an access class barring (ACB)
factor, alleviating pilot collision. Another extension can be found in \cite{SUCR-GBPA}, in which a graph-based pilot access (SUCR-GBPA) protocol is proposed, allowing all users that lost the
contention resolution to randomly select a new pilot. Besides, in \cite{Hebert_2021}, the SUCRe protocol is modified to include a non-orthogonal multiple access (NOMA) approach. The proposed NOMA-RA protocol allows to resolve collisions between users who try to access the medium using the same pilot signal, distinguishing them in the power domain and achieving a better sum-rate performance with lower average delay in comparison with the traditional SUCRe scheme.}

In the SUCRe protocol, the UE repeats its RA pilot in step~3 if it judges itself as the strongest contending UE, otherwise it stays silent. This can be seen as a hard decision, as described in \cite{SoftSUCRe}. As an improvement, the authors in~\cite{SoftSUCRe} derive the probability of each UE being the strongest contender, based on information available at the UE side. Then, in the proposed Soft-SUCRe protocol, the UE repeats or not its chosen RA pilot in step~3 according to this probability, achieving significant performance improvements in overcrowded scenarios, \emph{i.e.}, when the average number of accessing UEs is higher than the number of available RA pilots.

An important drawback of the original SUCRe protocol is its lack of fairness, since the strongest user retransmission criterion improves the connectivity performance of the UEs closer to the BS, while the farther UEs typically present increased probabilities of failed access attempts. In \cite{ACBPC}, the authors propose the \jcm{ACB} with power control (ACBPC) protocol, which is able to provide a good RA performance for all UEs, independently of their distances to the BS, improving the RA performance for a substantial fraction of UEs within the cell in comparison with SUCRe.

The SUCRe protocol has been recently adapted to the XL-MIMO scenario in \cite{Nishimura_2020}, in the so called SUCRe-XL. By allowing  different UEs to access the network through the same RA pilot, provided that the respective UE channels result in non-overlapping VRs {across the \jcm{XL} array of antennas}, this protocol achieves significant improvements in terms of reducing access latency and the fraction of failed access attempts in overcrowded scenarios. However, the SUCRe-XL RA protocol does not perform any specific action in the sense of favoring the communication of UEs with non-overlapping VRs channels by employing the same pilot, while expecting this to occur only at random.\\

\noindent\textit{Contribution}: This paper proposes a joint RA and scheduling \jcm{protocol} for crowded XL-MIMO systems, {named Non-Overlapping} VR XL-MIMO (NOVR-XL) {random access} protocol. \jcm{Different than other works regarding RA protocols, which terminate the analysis when the UE succeeds in the RA, we also look at the post-RA stage, in which payload data pilots are scheduled to the UEs that succeeded in the RA stage.} Due to the randomly distributed obstacles, scatterers, and UEs, different UEs might have different VRs {configurations}. In the RA stage, more than one UE may contend for the same RA pilot, giving rise to a pilot collision.  
Our protocol takes advantage of the random VR configuration of the different UEs to admit them to the network. {Once a new UE is admitted to the network, the proposed protocol tries to allocate a payload data pilot already in use by another UE, but with non-overlapping VRs.}  
In this way, the proposed 
{procedures} not only improve the RA {system} performance, 
but also increase the sum-rate, since more UEs 
{can be} connected to the same payload data pilot in an interference-free manner{, thanks to the large extension of the XL antenna array}. \jcm{Furthermore, another important advantage of our proposed NOVR-XL protocol is that it is \ta{just} a 2-step RA protocol, different than SUCRe-based ones which require \ta{four} steps, allowing thus an appreciable access latency reduction.}

\noindent{\it Notation}: The conjugate, transpose and conjugate-transpose of a matrix $\mathbf{A}$ are represented by $\mathbf{A}^*$, $\mathbf{A}^T$ and $\mathbf{A}^H$, respectively. The $i,j$-th element of a matrix $\bf A$ is $[{\bf A}]_{ij}$.  $\mathbf{I}_M$ is the $M\times M$ identity matrix, $\left| \cdot \right|$ and $\left\| \cdot \right\|$ represent the cardinality of a set and the Euclidean norm of a vector, respectively. In addition, $\mathcal{U} \backslash \mathcal{A}$ denotes the set containing elements that are in $\mathcal{U}$ but not in $\mathcal{A}$.  
$\mathcal{N}(\cdot,\cdot)$ denotes a Gaussian distribution, $\mathcal{CN}(\cdot,\cdot)$ is a circularly-symmetric complex Gaussian distribution, and $\mathcal{B}(\cdot,\cdot)$ is a binomial distribution. $\mathbb{C}$ 
denotes spaces of complex
-valued numbers, while $\Gamma(\cdot)$ is the Gamma function, and ${\rm Pr}(\cdot)$ denotes probability. The real part operator is $\Re(\cdot)$. $\mathbb{E}[\cdot]$ and ${\rm var}(\cdot)$ denote statistical expectation and variance, respectively.

\section{System model}\label{sec:model}

We consider a time-division-duplexing (TDD) system, in which a BS operates equipped with an extra-large uniform linear array (ULA) with $M$ antenna-elements. Similarly as in \cite{Wang20}, \cite{Nishimura_2020}, we consider a simplified bipartite graph model, being the array divided into $B$ SAs, each with $M_b = M/B$ antennas. Let we define $\mathcal{M}$ as the set of all the BS subarrays and $\mathcal{V}_k \subset \mathcal{M}$ the subset of SAs visible to $k$-th UE.
{For the sake of mathematical modeling, we assume that} the subset $\mathcal{V}_k$ is generated at random, where each SA follows a Bernoulli distribution with success probability $P_b$. This probability reproduces the effect of random obstacles and scatterers in the environment, which interact with the signals transmitted by/to the UEs, giving rise to the VRs. 
Thus, each SA is visible with probability $P_b$, allowing also a binary vector representation $\textbf{v}_k \in \mathbb{Z}^{B \times 1}$ for the VR of every UE, in which $v_{k,b} = 1$ if the $b$-th SA is visible by UE $k$, and 0 otherwise. 
{It is important to note that when an antenna is visible by some user, this means that the user can transmit/receive signals to/from this antenna with a non-zero channel gain, not necessarily implying in the existence of a line of sight (LOS) component.} 





\begin{figure}[htbp!]
\vspace{-5mm}
\centering
\includegraphics[trim={0mm 0mm 1mm 0mm},clip,width=.92\linewidth]{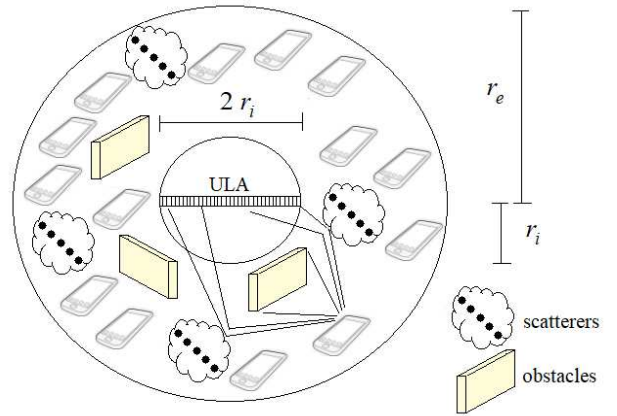}
\vspace{-5mm}
\caption{Illustration of the considered crowded scenario composed by an XL ULA and uniformly distributed UEs. The visibility region of a UE due to the presence of scatterers and obstacles is represented.} \label{fig: cell}
\end{figure}

Let $\mathcal{U}$ be the set of UEs in the cell, $\mathcal{A} \subset \mathcal{U}$ be the set of active UEs, with temporary dedicated data pilots, and $\mathcal{K} = \mathcal{U}\backslash\mathcal{A}$ be the set of the inactive UEs (iUEs), which do not have dedicated data pilots and need to be assigned one if they want to become active. 
Hence, $K = |\mathcal{K}|$ is the number of iUEs in the cell. \jcm{We assume a crowded scenario where $K$ iUEs ($K>M$) are uniformly distributed within a circular cell with internal and external radius of $r_i$ and $r_e$, respectively, as depicted in Fig. \ref{fig: cell}.}

We define $\tau_{\textsc{ra}}$ as the number of available {\it RA pilot sequences} $\boldsymbol{s}_1, \ldots,\boldsymbol{s}_{\tau_{\textsc{ra}}}\in \mathbb{C}^{\tau_{\textsc{ra}} \times 1}$. Each RA pilot has length $\tau_{\textsc{ra}}$ and $\lVert \boldsymbol{s}_t \rVert^2 = \tau_{\textsc{ra}}$, since we are considering mutually orthogonal pilot sequences. Similarly, $\tau_{\textsc{pd}}$ is the number of available {\it payload data pilot} (PDP) {\it sequences} $\boldsymbol{p}_1, \ldots,\boldsymbol{p}_{\tau_{\textsc{pd}}}\in \mathbb{C}^{\tau_{\textsc{pd}} \times 1}$, which also are orthogonal and have $\lVert \boldsymbol{p}_t \rVert^2 = \tau_{\textsc{pd}}$. We define the set of PDP sequences as $\mathcal{P}$, with cardinality $|\mathcal{P}| = \tau_{\textsc{pd}}$. In a practical system, some sequences in the set $\mathcal{P}$ would be temporarily allocated to the active UEs, and some other would be idle, being available to admit new UEs which succeed in the RA stage. However, in our model, we adopt the simplifying condition that all PDP sequences in $\mathcal{P}$ are temporarily allocated to the active UEs, while every UE succeeding in the RA stage can be admitted to the network.\footnote{In practice, it is like if the PDP length could be adjusted to match the number of PDPs allocated in that moment, or if the number of UEs disconnecting from the network matches the number of connecting UEs.} This is assumed to avoid the problem of finding a suitable PDP length $\tau_{\textsc{pd}} = |\mathcal{P}|$, trading-off the PDP transmission overhead and the availability of PDPs for suitable scheduling new UEs, which is beyond the scope of this paper.

The time-frequency resources are divided into coherence blocks of $T$ channel uses, dimensioned such that the channel responses between the BS and UEs are constant and frequency flat within each block, while varying between blocks. This {time-frequency} organization can be implemented with orthogonal frequency-division multiplexing (OFDM). Similarly as in \cite{Bjornson_2017}, the coherence blocks are divided into two categories: {\it payload data blocks} and {\it RA blocks}. The blocks in the first category are used for uplink (UL) and downlink (DL) payload data transmission by the users in the {subset} $\mathcal{A}$, which have been already allocated a temporary dedicated PDP {sequence} when they succeeded in an RA attempt. The second category is used for {RA} from iUEs in the set $\mathcal{K}$ when they wish to be admitted to the payload data blocks, which requires the {\it temporary} allocation of a PDP sequence. 

The large-scale fading {includes path loss and shadowing, following} an urban micro scenario model {given by \cite{Metis_2013}:}
\begin{equation}
 \beta^{(b)}_{k,m} = 10^{-\kappa \log(d^{(b)}_{k,m}) + \frac{g+\chi}{10}},
\end{equation}
where $d^{(b)}_{k,m}$ is the distance between {the $k$-th} UE and the $m$-th antenna ($m = 1,\ldots,M_b$) of SA $b$, $\chi \sim \mathcal{N}(0,\sigma^2_{\rm sf})$  is the shadow fading, with standard deviation $\sigma_{\rm sf} = 10$ dB; $\kappa = 3.8$ is the adopted path loss exponent, and  $g = - 34.53$ dB is the path loss at the reference distance. Similarly as in \cite{Nishimura_2020}, we adopt the simplification
 $\beta^{(b)}_k = \frac{1}{M_b}\sum_{m =1}^{M_b}\beta^{(b)}_{k,m}$, and blocked SAs resulting in $\beta^{(b)}_k = 0$.  Hence, $\beta^{(b)}_k$ represents the average large-scale fading coefficient for the $k$-th user at the $b$-th SA.

Let $\mathbf{h}^{(b)}_{k}\in \mathbb{C}^{M_b\times 1}$  be the Rayleigh fading channel vector  between UE $k\in \mathcal{K}$ and SA $b$, 
following $\mathbf{h}^{(b)}_{k}\sim \mathcal{CN}(0,\beta^{(b)}_k\mathbf{I}_{M_b})$. 
{The above {XL-MIMO channel} model assumes a non-line-of-sight (NLOS) channel between UE and BS, in the same way as \cite{Wang20, Ali_2020, Nishimura_2020, Marinello_20}. This assumption can be justified in practice for scenarios where a large number of obstacles and scatterers takes place between UEs and the BS.}


Under this model, as {shown} in \cite{Marinello_20}, the signal to interference plus noise ratio (SINR) of an active UE $k \in \mathcal{A}$ employing zero-forcing (ZF) combiner/precoder is {described by:} 
\begin{equation}\label{eq:SINR_ZF_as}
\overline{\gamma}_k^{(\textsc{ZF})} = \frac{\jcm{\overline{\rho}}}{\sigma^2} \left(M_b \, \sum_{b=1}^{B} \beta^{(b)}_k - \sum_{j \in \mathcal{A}, j\neq k} \frac{\sum_{b=1}^{B} \beta^{(b)}_k \beta^{(b)}_j}{\sum_{b=1}^{B} \beta^{(b)}_j} \right),
\end{equation}
in which $\jcm{\overline{\rho}}$ is the transmit power of \jcm{the UEs assuming a uniform power allocation policy in the data transmission stage}, and $\sigma^2$ is the noise variance. Besides, if this UE performs $\mu^{\textsc{ra}}_k$ RA attempts until establishing a connection, and then remains a number of  $\mu^{\textsc{pd}}_k$ channel coherence time ($t_c$) intervals transmitting payload data occupying the whole bandwidth {$\textsc{w}$}, its achieved spectral efficiency is given by:
%
\begin{equation}\label{eq:SE_ZF_as}
R_k = \left(\frac{\mu^{\textsc{pd}}_k {\textsc{w}} \, t_c }{\mu^{\textsc{ra}}_k \varphi_{\textsc{ra}} + \mu^{\textsc{pd}}_k {\textsc{w}} \, t_c}\right) \left(1 - \frac{\tau_{\textsc{pd}}}{T} \right) \log_2\left(1+\overline{\gamma}_k^{(\textsc{ZF})}\right),
\end{equation}
in which $\varphi_{\textsc{ra}}$ is the number of channel uses spent in each RA attempt, considering the employed RA protocol, and recalling that $T$ is the number of channel uses in a coherence time. It is noteworthy that the first pre-log factor in \eqref{eq:SE_ZF_as} accounts for the RA overhead, while the second one accounts for the PDPs transmission overhead for channel estimation purposes.

Finally, the system sum-rate {($\rm SR$)} \jcm{is given by}%
\begin{equation}\label{eq:SR_ZF_as}
{\rm SR} = {\textsc{w}} \cdot  \sum_{k \in \mathcal{A}} R_k.
\end{equation}
The sum-rate metric 
is very useful in order to compare different RA protocols, since it takes into account the number of channel uses spent per attempt, the number of RA attempts that the UEs are requiring to connect, and the number of active UEs that the protocol can admit to the network. Furthermore, since $\tau_{\textsc{pd}}$ always coincide in our model to the number of PDP temporarily allocated to the active UEs, and $\tau_{\textsc{pd}} \leq |\mathcal{A}|$ since the same PDP can be allocated to different active UEs if their VRs are non-overlapping, the sum-rate metric also quantifies the capability of the scheduling algorithm in finding UEs with orthogonal VRs to allocate the same PDP.

\section{Proposed NOVR-XL Protocol}\label{sec:RA_XL-MIMO_protocol}
In this section, we describe the two steps of our proposed NOVR-XL protocol to jointly perform RA and {PDP} scheduling in the XL-MIMO system. In the first step, all iUEs make an RA attempt with probability $P_a\leq 1$. Each iUE $k \in \mathcal{K}$ making an RA attempt uniformly selects an uplink RA pilot sequence $\boldsymbol{s}_{r(k)} \in \mathbb{C}^{\tau_{\textsc{ra}} \times 1}$, where $r(k) \in \{1,2,\ldots, \tau_{\textsc{ra}} \}$, transmitting it with
a non-zero power $\rho_k > 0$, otherwise it stays silent by setting $\rho_k = 0$. Since transmission is uncoordinated, it is possible and {utmost} usual {in crowded scenarios} that more than one iUE choose the same RA pilot sequence. Therefore, let $\mathcal{S}_t = \{ k : r(k) = t, \rho_k > 0\}$ represent the set of iUEs indices transmitting pilot $t$. As in \cite{Bjornson_2017}, its cardinality follows: 
\begin{equation}\label{eq: binomial distribution}
|\mathcal{S}_t|\sim \mathcal{B}\left(K,\frac{P_a}{\tau_{\textsc{ra}}}\right).
\end{equation}
where $\mathcal{B}(\cdot,\cdot)$ is the binomial distribution. 

\noindent\textit{\textbf{Step 1}: Random Pilot Sequence UL Transmission.} At this step, the inactive UEs 
{that want} to establish a connection 
{transmit} the signal $\textbf{x}_k = [\boldsymbol{s}_{r(k)}; \textbf{d}^{\rm ul}_k] \in \mathbb{C}^{(\tau_{\textsc{ra}}+n_u)\times 1}$, composed by its chosen UL RA pilot sequence and the additional information vector $\textbf{d}^{\rm ul}_k\in \mathbb{C}^{n_u\times 1}$. The additional information vector $\textbf{d}^{\rm ul}_k$ contains the identity of the UE, and its estimated VR binary vector $\widehat{\textbf{v}}_k \in \mathbb{Z}^{B \times 1}$. Our method assumes that $\widehat{\textbf{v}}_k$ can be obtained at the UE's side, employing the periodic transmission of a broadcast signal that activates each SA sequentially, which can be seen as a slight modification of the broadcast signal transmission in step~0, typically assumed by RA protocols, as in \cite{Bjornson_2017}, \cite{Nishimura_2020}. Thus, the $b$-th SA receives the signal
\begin{equation}\label{eq:step1: UE RA pilot1}
\textbf{Y}^{(b)} = [\textbf{Y}_p^{(b)}, \textbf{Y}_d^{(b)}] = \sum_{k\in \mathcal{K}} \sqrt{\rho_k}\mathbf{h}_k^{(b)} \textbf{x}_{k}^{T} + \textbf{N}^{(b)},
\end{equation}
where $\textbf{Y}^{(b)}\in  \mathbb{C}^{M_b\times(\tau_{\textsc{ra}}+n_u)}$, $\textbf{Y}_p^{(b)}\in  \mathbb{C}^{M_b\times\tau_{\textsc{ra}}}$, $\textbf{Y}_d^{(b)}\in  \mathbb{C}^{M_b\times n_u}$, and $\textbf{N}^{(b)}\in \mathbb{C}^{M_b\times (\tau_{\textsc{ra}}+n_u)}$ is the receiver noise with entries drawn from $\mathcal{CN}(0,\sigma^2)$. Besides, we have that
\begin{equation}\label{eq:step1: UE RA pilot2}
\textbf{Y}_p^{(b)} = \sum_{k\in \mathcal{K}} \sqrt{\rho_k}\mathbf{h}_k^{(b)} \boldsymbol{s}_{r(k)}^{T} + \textbf{N}_p^{(b)}, \,\, {\rm and}
\end{equation}
\begin{equation}\label{eq:step1: UE RA pilot3}
\textbf{Y}_d^{(b)} = \sum_{k\in \mathcal{K}} \sqrt{\rho_k}\mathbf{h}_k^{(b)} (\textbf{d}^{\rm ul}_{k})^{T} + \textbf{N}_d^{(b)},
\end{equation}
with $\textbf{N}^{(b)} = [\textbf{N}_p^{(b)}, \textbf{N}_d^{(b)}]$.
{Hence,} each SA correlates \eqref{eq:step1: UE RA pilot2} 
{with each RA pilot. For the case of an arbitrary RA pilot $\boldsymbol{s}_t$, it yields:} 
\begin{equation}\label{eq:step1_yt}
\textbf{y}_t^{(b)} = \textbf{Y}_p^{(b)}\frac{\boldsymbol{s}_t^*}{\lVert \boldsymbol{s}_t \rVert} = \sum_{i\in \mathcal{S}_t} \sqrt{\rho_i \tau_{\textsc{ra}}}\mathbf{h}_i^{(b)}  + \mathbf{n}^{(b)}_t, \quad b=1,\ldots, B,
\end{equation} 
where $\mathbf{n}^{(b)}_t = \textbf{N}_p^{(b)}\frac{\boldsymbol{s}_t^*}{\lVert \boldsymbol{s}_t \rVert}$ is the effective receiver noise, so that $\mathbf{n}^{(b)}_t \sim \mathcal{CN}(0,\sigma^2 \mathbf{I}_{M_b})$. {As a result}, each SA tries to decode the message contained in \eqref{eq:step1: UE RA pilot3} using the estimates $\textbf{y}_t^{(b)}$, evaluating\footnote{Note that, at this moment, the BS does not know which RA pilot each UE has chosen. So, the BS tries do decode the message in \eqref{eq:step1: UE RA pilot3} using the estimates obtained from every RA pilot. Only when the decoding is successful the BS identifies which UE transmitted that message.}:
\begin{align}\label{eq:d_hat_gen}
\widehat{\textbf{d}^{\rm ul}_{k}}^T = \frac{(\textbf{y}_{t}^{(b)})^H}{\sqrt{\tau_{\textsc{ra}}}} \textbf{Y}_d^{(b)}.
\end{align}

In the next section, we {analyze} the attainable SINR of {$\widehat{\textbf{d}^{\rm ul}_{k}}$} in \eqref{eq:d_hat_gen}, while pointing out the conditions in which its information can be suitably decoded. This is very likely to occur when the SA $b$ is {\it exclusively} visible to the {$k$-th} UE among the UEs contending by RA pilot {$t= s_{r(k)}$, \emph{i.e.}, in $\mathcal{S}_{t}$}. This means that $b \in \mathcal{V}_k$ and $b \not\in \mathcal{V}_{k'}$, $\forall k' \in \mathcal{S}_t, k' \neq k$. In this case, $\textbf{y}_{t}^{(b)} = \sqrt{\rho_{k} \tau_{\textsc{ra}}}\mathbf{h}_{k}^{(b)}  + \mathbf{n}^{(b)}_{t}$, {representing} a very good channel estimate of $\mathbf{h}_{k}^{(b)}$. However, there is a possibility that the $b$-th SA may be visible to more than one UE in $\mathcal{S}_t$, with the $k$-th UE presenting a strong predominant channel gain w.r.t. its contenders, so that the channel estimate in \eqref{eq:step1_yt}, although partially contaminated, {could be} of sufficient quality to result in an appropriate decoding of the additional information from this UE. All these possibilities are considered in the SINR analysis conducted in the next section. 
%
%

{Assuming} from now on an appropriate decoding of the information in \eqref{eq:d_hat_gen}, the UE 
$k$ is admitted to the set $\mathcal{C}$, which contains the indices of UEs that wished to become active and had their additional information in  \eqref{eq:d_hat_gen} suitably decoded in some SA. 
Besides, the BS gets to know the {estimated} VR of UEs in the set $\mathcal{C}$, constructing the binary matrix $\textbf{V} \in \mathbb{Z}^{B \times |\mathcal{C}|}$, composed by the VR vectors $\textbf{v}_k$ of the UE $k \in \mathcal{C}$. It also creates the binary matrix $\textbf{E} \in \mathbb{Z}^{B \times |\mathcal{C}|}$, whose $i,j${-th} element is set when the BS has {\it reliable channel estimates} in SA $i$ for the $j$-th UE in $\mathcal{C}$. This occurs when the BS successfully decoded the message $\textbf{d}^{\rm ul}_{j}$ from {the $i$-th} SA. 
In such cases, the BS stores the channel estimation vectors $\widehat{\bf h}^{(b)}_{k} = \widehat{\bf y}^{(b)}_{r(k)}$, when $[\textbf{E}]_{b,k} = 1$; otherwise, $\widehat{\bf h}^{(b)}_k = {\bf 0}_{M_b}$ 
{and $[\textbf{E}]_{b,k} = 0$}. 

Conventionally, in a spatially stationary channel configuration, it is necessary a number of $|\mathcal{C}|$  payload data pilot (PDP) {sequences} be available to admit the UEs in $\mathcal{C}$ into the active set $\mathcal{A}$. Due to the high density of UEs in crowded XL-MIMO scenarios and scarcity of orthogonal PDP sequences, such exclusive allocation of PDPs to active UEs would require 
{sequences} of long lengths, substantially increasing the training overhead and spectral efficiency penalty in \eqref{eq:SE_ZF_as}. 
However, due to the obstacles and scatterers in {typical wireless environment}, combined with the {XL-MIMO propagation features}, the spatial non-stationarities and distinct VRs per UE emerge; hence, it is 
reasonable and usual that some UEs in $\mathcal{C}$ present {\it non-overlapping} VRs {with respect to} certain active UEs in $\mathcal{A}$. Moreover, since the BS already knows their VRs, this implies that the same PDP sequence can be allocated to such UEs, since they can communicate with the same data pilot in an interference-free manner, due to the orthogonality in the VR-{spatial} domain.

We propose a {novel} procedure for the BS to allocate PDP sequences. First, the BS needs to update matrix ${\bf F} \in \mathbb{Z}^{|\mathcal{P}|\times B}$, which contains the combined VRs of UEs allocated to each PDP sequence. The combined VRs related to each allocated PDP is simply the sum of the VRs of the UEs using it. Since the system starts to operate, for each UE $k$ succeeding in the RA stage, the BS first seeks for a line in $\bf F$ that is orthogonal to ${\bf v}_k$
{. By denoting the $j$-th line of ${\bf F}$ by ${\bf f}_j$, this operation is equivalent to finding ${\bf f}_{j} \, {\bf v}_k = 0$.} Once this condition is valid for some line $j$, the $j$-th PDP in $\mathcal{P}$ can be allocated to the UE $k$. In this case, the matrix $\bf F$ is updated, evaluating 
{${\bf f}_{j} \leftarrow {\bf f}_{j} + {\bf v}^T_k$.} On the other hand, if the VR of the UE $k$ is not orthogonal to the combined VR of any PDP already allocated, the BS has to allocate a new PDP {sequence} to this UE. In this case, ${\bf F} {\leftarrow} \left[ {\bf F}^T, {\bf v}_k \right]^T$, and $\mathcal{P} {\leftarrow} \mathcal{P} \cup \boldsymbol{p}_t$, being $\boldsymbol{p}_t$ the new PDP allocated to the UE $k$. Furthermore, whenever the UE $k$ which was allocated the PDP {sequence} $j$ terminates its connection, matrix $\bf F$ is updated evaluating 
{${\bf f}_{j} \leftarrow {\bf f}_{j} - {\bf v}^T_k$.} If this line becomes null, it is removed from matrix $\bf F$, as well as set $\mathcal{P}$ is updated as $\mathcal{P} {\leftarrow} \mathcal{P} \backslash \boldsymbol{p}_j$. Our proposed {PDP} scheduling procedure is presented in Algorithm \ref{alg:sched}.

\begin{algorithm}[!htbp]
\caption{\small{\jcm{NOVR-XL} Scheduling Procedure}}
\begin{flushleft}
Input: {$\mathcal{P}$, $\textbf{F}$, ${\bf v}_k$.}
\end{flushleft}
\label{alg:sched}
\begin{algorithmic}[1]
\STATE{Initialize the index of the PDP to be allocated to UE $k$ as $\iota {\leftarrow} -1$;}
\FOR{$j = 1, 2, \ldots, |\mathcal{P}|$}
    \IF{{${\bf f}_{j} \, {\bf v}_k == 0$}}
        \STATE{Evaluate $\iota {\leftarrow} j$;}
        \STATE{Evaluate {${\bf f}_{j} \leftarrow {\bf f}_{j} + {\bf v}^T_k$};}
        \STATE{\textbf{break};}
    \ENDIF
\ENDFOR	
\IF{$\iota == -1$}
    \STATE{Obtain a new PDP $\boldsymbol{p}_t$ to be allocated to the UE $k$;}
    \STATE{Evaluate ${\bf F} {\leftarrow} \left[ {\bf F}^T, {\bf v}_k \right]^T$;}
    \STATE{Evaluate $\mathcal{P} {\leftarrow} \mathcal{P} \cup \boldsymbol{p}_t$;}
    \STATE{Evaluate $\iota {\leftarrow} t$;}
\ENDIF
\end{algorithmic}
Output: $\mathcal{P}$, $\textbf{F}$, $\iota$.
\end{algorithm}

\vspace{2mm}

%
\noindent\textit{\textbf{Step 2}: Precoded Random Access DL Response.}
The $b$-th SA generates a precoded DL signal
\begin{equation}\label{eq:step2: DL response}
\mathbf{G}^{(b)} = \sqrt{\frac{q}{B\, \delta_b}}\sum_{k \in \mathcal{C}} [{\bf E}]_{b,k}  \frac{\widehat{\mathbf{h}}_k^{(b)*}}{\lVert \widehat{\mathbf{h}}_k^{(b)} \rVert}(\textbf{d}^{\rm dl}_{k})^T,
\end{equation}
where the DL precoder matrix $\mathbf{G}^{(b)}\in \mathbb{C}^{M_b\times n_d}$, $q$ is the BS transmit power, $\delta_b = \sum_{k \in \mathcal{C}} [{\bf E}]_{b,k}$ is the number of UEs in $\mathcal{C}$ having reliable channel estimates in SA $b$, and $\textbf{d}^{\rm dl}_{k} \in \mathbb{C}^{n_d \times 1}$ is the DL RA data destined to the $k$-th UE, containing the index of the PDP allocated to it, according to the output of Algorithm \ref{alg:sched}, and other information, like timing advance, etc. Note that in \eqref{eq:step2: DL response} the DL data of a given UE is precoded only in the SAs where the BS has reliable channel estimates, since $[{\bf E}]_{b,k}=0$ in the other SAs. Besides, it is not necessary to send DL pilots at this step of {the} proposed {NOVR-XL} protocol, differently than SUCRe-XL, since the UEs will not estimate the sum of contenders signal gains 
{in this new context}. 

Thus, the UE ${k} \in \mathcal{C}$ receives, in the reciprocal channel, the signal 
{$\widehat{\textbf{d}^{\rm dl}_{k}}^T \in \mathbb{C}^{1 \times n_d}$},  
%
\begin{align}\label{eq:step2:rx_signal_UE}
\widehat{\textbf{d}^{\rm dl}_{k}}^T &= \sum_{b \in \mathcal{V}_{k}}\mathbf{h}_{k}^{(b)T} \mathbf{G}^{(b)} + \boldsymbol{\eta}_{k}^T,\\
&= \sum_{b \in \mathcal{V}_{k}} \sqrt{\frac{q}{B\, \delta_b}} [{\bf E}]_{b,k} \mathbf{h}_{k}^{(b)T}  \frac{\widehat{\mathbf{h}}_{k}^{(b)*}}{\lVert \widehat{\mathbf{h}}_{k}^{(b)} \rVert}(\textbf{d}^{\rm dl}_{k})^T + \ldots \notag\\ 
& + \sum_{b \in \mathcal{V}_{k}} \sqrt{\frac{q}{B\, \delta_b}} \mathbf{h}_{k}^{(b)T} \sum_{j \in \mathcal{C}, j\neq k} [{\bf E}]_{b,j}  \frac{\widehat{\mathbf{h}}_{j}^{(b)*}}{\lVert \widehat{\mathbf{h}}_{j}^{(b)} \rVert}(\textbf{d}^{\rm dl}_{j})^T \notag  + \boldsymbol{\eta}_{k}^T,
\end{align}where {$\boldsymbol{\eta}_{k} \sim \mathcal{CN}(0,\sigma^2 \textbf{I}_{n_d})$} is noise. 
\normalsize
%
%
In the next section, the SINR of $\widehat{\textbf{d}^{\rm dl}_{k}}$ in \eqref{eq:step2:rx_signal_UE} is evaluated, pointing out the conditions in which it can be suitably decoded at the UE. 

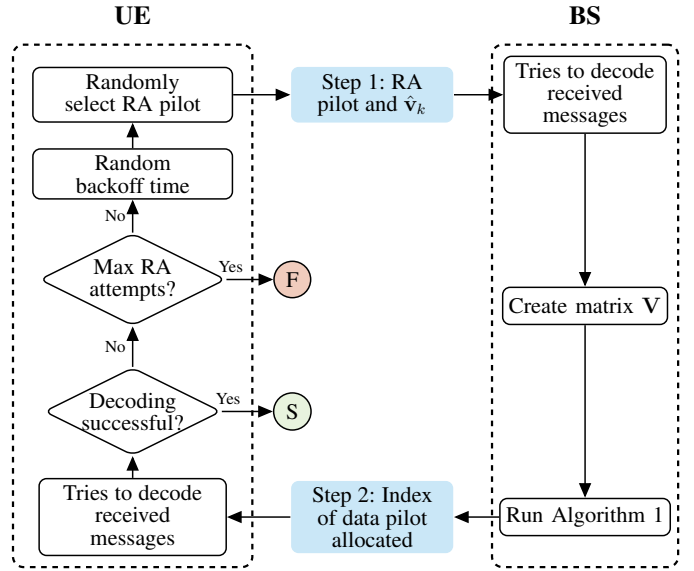
\begin{figure}[htbp!]
\vspace{-2mm}
\centering
\scalebox{0.7}{
	\begin{tikzpicture}[auto, thick, node distance=2cm, >=triangle 45]
		\node[block, text width=10em, text centered] (RApilot) at (0,0) {\large Randomly select RA pilot};
		\node[block, rounded corners, minimum size=3em, fill=Cerulean!20, draw=none, text width=8em, text centered] (step1) at (4.5,0) {\large Step 1: RA pilot and $\hat{\bf v}_k$};
		\node[block, minimum width=8em, text width=8em, text centered] (BS1) at (8.5,0) {\large Tries to decode received messages};
		\node[block, minimum width=8em] (BS2) at (8.5,-4) {\large Create matrix $\bf V$};
		\node[block, minimum width=8em] (BS3) at (8.5,-8) {\large Run Algorithm 1};
		\node[block, rounded corners, minimum size=3em, fill=Cerulean!20, draw=none, text width=8em, text centered] (step2) at (4.5,-8) {\large Step 2: Index of data pilot allocated};
		\node[block, text width=8em, text centered] (UE1) at (0,-8) {\large Tries to decode received messages};
		\node[decision, inner sep=-1pt, text width=6em, text centered] (UE2) at (0,-6) {\large Decoding successful?};
		\node[decision, inner sep=-1pt, text width=6em, text centered] (UE3) at (0,-3.5) {\large Max RA attempts?};
		\node[block, text width=10em, text centered] (UE4) at (0,-1.5) {\large Random backoff time};
		\node[circulo, fill=BrickRed!20] (Failure) at (3,-3.5) {\large F};
		\node[circulo, fill=YellowGreen!20] (Success) at (3,-6) {\large S};
				
		\node (A) at (-2,0.7) {};
		\node (B) at (2,-8.7) {};
		\node (C) at (7,0.7) {};
		\node (D) at (10,-8.7) {};
		\node (UE) at (0,1.5) {\Large {\bf UE}};
		\node (BS) at (8.5,1.5) {\Large {\bf BS}};
				
		\draw[->](RApilot) -- (step1);
		\draw[->](step1) -- (BS1);
		\draw[->](BS1) -- (BS2);
		\draw[->](BS2) -- (BS3);
		\draw[->](BS3) -- (step2);
		\draw[->](step2) -- (UE1);
		\draw[->](UE1) -- (UE2);
		\draw[->](UE2) -- node{\small No}(UE3);
		\draw[->](UE3) -- node{\small No}(UE4);
		\draw[->](UE4) -- (RApilot);
				
		\draw[->, above left](UE2) -- node{\small Yes}(Success);
		\draw[->, above left](UE3) -- node{\small Yes}(Failure);
				
		\begin{pgfonlayer}{background} 
			\node[surround, fill=none, dashed, very thick] (background) [fit = (A) (B)] {};
			\node[surround, fill=none, dashed, very thick] (background) [fit = (C) (D)] {};
		\end{pgfonlayer}
	\end{tikzpicture}
}
\vspace{-2mm}
\caption{Diagram of the proposed procedures for joint RA and {PDP} scheduling in XL-MIMO systems. "F" denotes a failed access attempt, while "S" denotes a successful access attempt.}	
\label{fig:flux}
\end{figure}

At this point, all UEs that 
wished to connect to the network try do decode the received signal. If the decoding is successful, this UE {will complete the admission process} to the active users set $\mathcal{A}$, being temporarily allocated a PDP sequence from $\mathcal{P}$, while achieving a spectral efficiency given by \eqref{eq:SE_ZF_as}. If the decoding fails, this UE performs another RA attempt after a random backoff time, as in \cite{Bjornson_2017}. If the UE does not succeed in connecting to the network after {a predefined number of} {RA} attempts, it gives up and declares a failed access attempt \cite{Bjornson_2017}.
Fig. \ref{fig:flux} summarizes the main steps of the proposed {NOVR-XL RA} procedure, highlighting the two steps of the {RA} protocol, and the main actions performed by the UE and BS.


\subsection{Variant of SUCRe-XL RA protocol: Re-allocating PDPs due to Orthogonal VRs} \label{sec:variant_RA_XL-MIMO}
Besides considering the SUCRe-XL RA protocol as proposed in \cite{Nishimura_2020}, for performance comparison in the next section, we also consider a variant of this protocol denominated {\it modified} SUCRe-XL (mSUCRe-XL) RA protocol, with the aim of leveraging non-overlapping VRs {features}. In this variant, the four steps of the SUCRe-XL protocol are performed similarly as in \cite{Nishimura_2020}, except that, in step~3, the re-transmitting UEs transmit together with their repeated RA pilot the additional information vector containing ${\bf \widehat{v}}_k$, estimated from the modified broadcast signal in step~0. If the collision is successfully resolved, the BS can decode such information and gets to know the VR of this new UE. Then, the BS is able to run our proposed scheduling procedure described in Algorithm \ref{alg:sched} in order to seek a PDP {sequence} allocated to active UEs whose combined VRs are orthogonal to that of UE $k$. 
This variant serves as a challenging opponent to the proposed NOVR-XL RA protocol.

In summary, the original SUCRe-XL protocol from \cite{Nishimura_2020} does not provide to the BS an explicit information about the VRs of the UEs. Therefore, each UE admitted to the network has to be temporarily allocated an exclusive PDP sequence to it, such that $|\mathcal{P}| = |\mathcal{A}|$. On the other hand, in the mSUCRe-XL protocol discussed herein, little overhead is added in the step~3 of the RA protocol, aiming to allow a more efficient scheduling of PDP resources, such that $|\mathcal{P}| \leq |\mathcal{A}|$. Moreover, the sum-rate metric in \eqref{eq:SR_ZF_as} is numerically evaluated in the next section, and the presented results can answer if the re-allocation of PDPs to UEs with non-overlapping VRs {compensates} 
the additional RA overhead required.

\jcm{Finally, it is \ta{worth mentioning} that SUCRe-XL and mSUCRe-XL RA protocols require \ta{four} steps since they rely on the \emph{strongest user} retransmission criterion, \emph{i.e.}, only one UE (with the strongest channel gain) retransmits the RA pilot in step 3, resolving the collision. For this purpose, it is required the DL pilot transmission in step 2, allowing the UEs to estimate the sum of contenders' signal gains, and then judging if it is the strongest. This approach is not effective from the connectivity perspective, since its objective is to connect only one UE per RA pilot\ta{; also, it} is unfair, since such UE is the one closer to the BS, \ta{while} increases access latency, since \ta{four} steps are required. Contrarily, our proposed NOVR-XL protocol does not rely on the strongest user retransmission criterion\ta{; hence, } 
the RA attempt for a UE \ta{will be} successful if their RA message can be decoded from at least one SA, which is likely to occur due to the different random VRs of the UEs in the XL-MIMO system. 
\ta{As a result,} more UEs can connect through the same RA pilot \ta{resource}, improving RA performance 
\ta{in just} 2 steps, 
since no DL pilot transmission is required, reducing access latency.}



\section{\ta{SINR in the NOVR-XL Protocol}} 
\ta{This section develops analytical expressions for the uplink and downlink SINR of the proposed NOVR-XL protocol. We start analyzing the probability of a specific UE channel generates a visible region under an exclusive SA. Hence, the expressions for the SINR in step 1 (UL) and step 2 (DL) of the proposed algorithm are obtained analytically.}
\label{sec:sinr}

\subsection{Probability of a UE having an exclusive SA} 
We first evaluate the probability of an SA $b$ being exclusively visible by UE $k$, among the UEs contending for the same RA pilot. In this case, 
$b \in \mathcal{V}_k$ and $b \not\in \mathcal{V}_{k'}$, $\forall k' \in \mathcal{S}_t, k' \neq k$. This condition is very desirable for our proposed NOVR-XL protocol, because it provides reliable channel estimates in that SA, and, consequently, reliable communication links for that UE in both directions. This makes very likely that the BS will properly decode the information message sent by the UE in step~1, besides of allowing the BS to transmit in step~2 information about the PDP being allocated for it. However, it is important to note that even if no SA is exclusively visible by UE $k$ among its contenders, there is still a chance for its information message sent in step~1 being successfully decodable. This is the case when the $k$-th UE presents a strong predominant channel gain in certain SA $b$ w.r.t. its contenders, so that the channel estimate in \eqref{eq:step1_yt} for this SA, although partially contaminated, is of sufficient quality for an appropriate decoding of the additional information from this UE. All these possibilities are discussed in the SINR analysis conducted in {Subsection \ref{SINR-UL-stp1}.}

{From the} system {model 
in} Section \ref{sec:model}, the probability of SA $b$ being visible by UE $k$ is $P_b$; therefore, given a number of UEs $|\mathcal{S}_t|$ contending for the same RA pilot, the probability of the SA $b$ being exclusively visible to it 
{can be written as} 
\begin{equation}
{P^{\textsc{sa}}_{\rm exc}} = P_b \cdot \left( 1-P_b \right)^{|\mathcal{S}_t|-1}.
\end{equation}

Since we need that at least one SA is exclusively visible by the UE, this probability $P_{\rm exc}$ can be evaluated as
\begin{equation}
    P_{\rm exc} = 1 - \left( 1-P^{{\textsc{sa}}}_{\rm exc} \right)^B.
\end{equation}

\begin{figure}[htbp!]
\vspace{-2mm}
\centering
\includegraphics[trim={1mm 1mm 1mm 0mm},clip,width=1\linewidth]{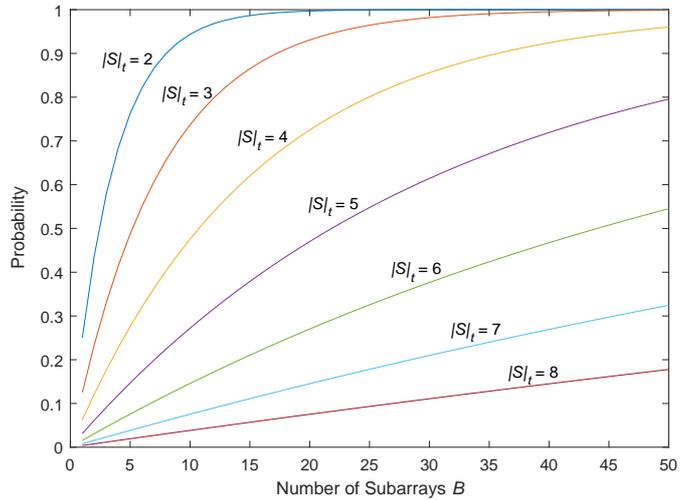}
\vspace{-5mm}
\caption{Probability of a UE having at least one SA exclusively visible by it, for $P_b = 0.5$. If $P_b$ increases, the curves move towards the bottom right corner, while if $P_b$ decreases, the curves move towards the upper left corner.}
\label{fig:PSA_exc}
\end{figure}

{Fig.} \ref{fig:PSA_exc} depicts the probability of the UE having at least one SA exclusively visible by it. As one can see, such probability achieves appreciable values for a not so large number of UEs contending for the same RA pilot. This corroborates the idea of the proposed NOVR-XL protocol in taking advantage of the reliable channel links in the RA stage for increasing the number of UEs admitted to the network.

\subsection{SINR of the UL Information in Step 1}\label{SINR-UL-stp1}
We derive herein the SINR performance {for the UL data} $\widehat{\textbf{d}^{\rm ul}_{k}}$ as {defined in} \eqref{eq:d_hat_gen}. First, we expand the components of the $\ell$-th term ($\ell = 1,\ldots,n_u$) of this estimation in \eqref{eq:d_hat_gen2}. Then, following the analysis in \cite{Marinello17}, we compute the expected power of each term in \eqref{eq:d_hat_gen2} as the expectation of its squared magnitude. This result is shown in Table \ref{tab:power}, in terms of the squared mean and variance of each component. Then, taking only $\mathbb{E}[|\varpi_1|]^2$ as the desired signal power, while relegating $\ta{\mathbb{V}}{\rm ar}[|\varpi_1|]$ as an additional interference as in \cite{Marinello17}, and after some manipulations, we obtain the UL SINR  in \eqref{eq:SINR_full}. \jcm{One can recognize the first term in the denominator as the directed interference in the case of a contending UE also having the \ta{$b$-th SA visible}, which can be seen as a \ta{\it coherent} RA \ta{\it interference}. The second term can thus be seen as a \ta{\it non-coherent} RA \ta{\it interference}, and relates to the product of the variance of the channel estimate with the variance of the received signal at the \ta{$b$-th} SA in step 1.}
\begin{figure*}
\begin{align}\label{eq:d_hat_gen2}
\widehat{d^{\rm ul}_{k\ell}} &= \frac{(\textbf{y}_{t}^{(b)})^H}{\sqrt{\tau_{\textsc{ra}}}} [\textbf{Y}_d^{(b)}]_{:\ell},\nonumber\\
&= \frac{1}{\sqrt{\tau_{\textsc{ra}}}}\left[ \sum_{i\in \mathcal{S}_t} \sqrt{\rho_i \tau_{\textsc{ra}}}\mathbf{h}_i^{(b)}  + \mathbf{n}^{(b)}_t\right]^H \left[ \sum_{j\in \mathcal{K}} \sqrt{\rho_j}\mathbf{h}_j^{(b)} d^{\rm ul}_{j \ell} + [\textbf{N}_d^{(b)}]_{: \ell} \right],\nonumber\\
&= \underbrace{\rho_k ||\mathbf{h}_k^{(b)}||^2 d^{\rm ul}_{k \ell}}_{\varpi_1} + \underbrace{\sum_{i\in \mathcal{S}_t, i\neq k} \rho_i ||\mathbf{h}_i^{(b)}||^2 d^{\rm ul}_{i \ell}}_{\varpi_2} + \underbrace{\sum_{i \in \mathcal{S}_t} \sum_{j\in \mathcal{K}, j \neq i} \sqrt{\rho_i \rho_j } (\mathbf{h}_i^{(b)})^H \mathbf{h}_j^{(b)} d^{\rm ul}_{j \ell}}_{\varpi_3} + \underbrace{\frac{(\mathbf{n}^{(b)}_t)^H}{\sqrt{\tau_{\textsc{ra}}}} \sum_{j\in \mathcal{S}_t} \sqrt{\rho_j}\mathbf{h}_j^{(b)} d^{\rm ul}_{j \ell}}_{\varpi_4} +\notag\\
& + \underbrace{\frac{(\textbf{y}_{t}^{(b)})^H}{\sqrt{\tau_{\textsc{ra}}}} \sum_{j \in \mathcal{K}\backslash \mathcal{S}_t} \sqrt{\rho_j}\mathbf{h}_j^{(b)} d^{\rm ul}_{j \ell}}_{\varpi_5} + \underbrace{\frac{(\textbf{y}_{t}^{(b)})^H}{\sqrt{\tau_{\textsc{ra}}}} [\textbf{N}_d^{(b)}]_{: \ell}}_{\varpi_6}.
\end{align}
\normalsize
\end{figure*}

\begin{table*}[!htbp]
\caption{Expected power of the terms in \eqref{eq:d_hat_gen2}.}
\vspace{-2mm}
\centering
\small
\begin{tabular}{|c|c|c|}
\hline
\bf Term $\varpi_i$ & $\mathbb{E}[|\varpi_i|]^2$ & \bf $\ta{\mathbb{V}{\rm ar}}[|\varpi_i|]$\\
\hline
\hline
$\varpi_1$ & $M^2_b \, \rho^2_k \, (\beta_k^{(b)})^2$ & $M_b \, \rho_k \, \beta_k^{(b)}$\\
\hline
$\varpi_2$ & $M^2_b \sum_{\substack{i\in \mathcal{S}_t \\ i \neq k}} \rho^2_i \, (\beta_i^{(b)})^2$ & $M_b \sum_{\substack{i\in \mathcal{S}_t \\ i \neq k}} \rho_i \, \beta_i^{(b)}$\\
\hline
$\varpi_3$ & 0 & $M_b \sum_{i \in \mathcal{S}_t} \sum_{j\in \mathcal{K}, j \neq i} \rho_i \rho_j  \beta_i^{(b)} \beta_j^{(b)}$\\
\hline
$\varpi_4$ & 0 & $\frac{M_b \sigma^2}{\tau_{\textsc{ra}}} \sum_{j\in \mathcal{S}_t} \rho_j \beta_j^{(b)}$\\
\hline
$\varpi_5$ & 0 & $ M_b \left[\sum_{i\in \mathcal{S}_t} \rho_{i} \beta_{i}^{(b)} + \frac{\sigma^2}{\tau_{\textsc{ra}}} \right] \sum_{j \in \mathcal{K} \backslash \mathcal{S}_t} \rho_{j} \beta_{j}^{(b)}$\\
\hline
$\varpi_6$ & 0 & $ M_b \left[\sum_{i\in \mathcal{S}_t} \rho_{i} \beta_{i}^{(b)} + \frac{\sigma^2}{\tau_{\textsc{ra}}} \right] \sigma^2$\\
\hline
\end{tabular}
\label{tab:power}
\end{table*}
\normalsize

\begin{figure*}
\begin{equation}\label{eq:SINR_full}
\gamma^{{\rm ul}}_{b,k} = \frac{M_b \, \rho^2_k \, (\beta_k^{(b)})^2}{M_b \sum_{\substack{i\in \mathcal{S}_t \\ i \neq k}} \rho^2_{i} \, (\beta_{i}^{(b)})^2 + \left[\sum_{i\in \mathcal{S}_t} \rho_{i} \beta_{i}^{(b)} + \frac{\sigma^2}{\tau_{\textsc{ra}}} \right] \left[\sum_{j \in \mathcal{K}}\rho_{j} \beta_{j}^{(b)}+\sigma^2 \right]}.
\end{equation}
\end{figure*}

In the cases that SA $b$ is exclusively visible by UE $k$ among its contenders in $\mathcal{S}_t$, as described in the previous subsection, the SINR performance in \eqref{eq:SINR_full} can be simplified to
\begin{equation}\label{eq:SINR-UL-stp1}
\gamma^{{\rm ul}}_{b,k} = \frac{M_b \, \rho^2_k \, (\beta_k^{(b)})^2}{\left[\rho_k \beta_k^{(b)} + \frac{\sigma^2}{\tau_{\textsc{ra}}} \right] \left[\sum_{j \in \mathcal{K}}\rho_{j} \beta_{j}^{(b)}+\sigma^2 \right]}.
\end{equation}

\subsection{SINR of the DL Information in Step 2}
In order to derive the SINR of the DL received signal at the UE in step~2, as described in \eqref{eq:step2:rx_signal_UE}, we need first to define the subset $\mathcal{I}_k^{(b)}$. This subset contains the indices of the UEs that result in a directed DL interference from SA $b$ to the UE $k$, \emph{i.e.}, if $k' \in \mathcal{I}^{(b)}_k$, the DL signal from SA $b$ is intended to UE $k'$, but it is also partially beamformed to UE $k$. Although rare, this might occur if the following four conditions simultaneously hold: \textbf{\textit{i}}) UEs $k$ and $k'$ chose the same RA pilot in step~1: $r(k) = r(k')$; \textbf{\textit{ii}}) SA $b$ is visible by UE $k$: $b \in \mathcal{V}_k$; \textbf{\textit{iii}}) SA $b$ is visible by UE $k'$: $b \in \mathcal{V}_k'$; and \textbf{\textit{iv}}) the BS acquired reliable channel estimates for UE $k'$ from SA $b$: $[{\bf E}]_{b,k'}=1$. This last condition refers to the cases in which SA $b$ is not exclusively visible by UE $k'$ among the UEs in the subset $\mathcal{S}_{r(k')}$, but it is also visible by UE $k$. However, since $\beta^{(b)}_{k'} \gg \beta^{(b)}_{k}$, the UL information from UE $k'$ in step~1 could be successfully decoded from SA $b$. Furthermore, note that condition \textbf{\textit{iv}} implies in condition \textbf{\textit{iii}}, since an SA can only obtain reliable channel estimates for a UE if this UE is visible by that SA.

We also define the constants $\alpha_t^{(b)}$ {assuming that the channel hardening condition holds:}
\begin{equation}
    \left(\alpha_t^{(b)}\right)^2 = \frac{||{\bf y}_t^{(b)}||^2}{{M_b}} \xrightarrow{M_b\rightarrow\infty} \sum_{i\in \mathcal{S}_t} \rho_i \tau_{\textsc{ra}} \beta_i^{(b)} + \sigma^2.
\end{equation}

Then, following a similar analysis conducted in Subsection~{\ref{SINR-UL-stp1}}, the SINR of the DL signal received at the UE $k$ in step~2 {can be} obtained as in \eqref{eq:SINR_DL_full}. One can note that the first term in the denominator refers to the directed DL interference due to the UEs in $\mathcal{I}_k^{(b)}$. Besides, the second term refers to conventional DL interference due to the other UEs in $\mathcal{C}$, and the third term refers to the noise power.\\

\begin{figure*}
\begin{equation}\label{eq:SINR_DL_full}
    \gamma^{{\rm dl}}_{k} = \frac{M_b \sum_{b \in \mathcal{V}_k} [{\bf E}]_{b,k} \frac{q \, \rho_k \, \tau_{\textsc{ra}}}{B \, \delta_b} \frac{(\beta_k^{(b)})^2}{(\alpha_{r(k)}^{(b)})^2}}{M_b \sum_{b \in \mathcal{V}_k}\sum_{k' \in \mathcal{I}^{(b)}_k} \frac{q \, \rho_k \, \tau_{\textsc{ra}}}{B \, \delta_b} \frac{(\beta_k^{(b)})^2}{(\alpha_{r(k')}^{(b)})^2} + \sum_{b \in \mathcal{V}_k}\sum_{k' \in \mathcal{C}} [{\bf E}]_{b,k'} \frac{q}{B \, \delta_b} \frac{\beta_k^{(b)}}{\alpha_{r(k')}^{(b)}} + \sigma^2}.
\end{equation}
\end{figure*}

\section{Numerical results} 
\label{sec:results}

\jcm{We assume a circular cell as depicted in Fig. \ref{fig: cell}, with $r_e = 200$m and $r_i = 20$m.} Each UE wants to become active with probability $P_a = 0.01$, randomly choosing one RA pilot out of $\tau_{\textsc{ra}} = 10$ {pilots} available. The BS is a ULA located in the center of the cell where $M = 400$ antennas are uniformly distributed along a $40$ m length array, and divided into $B$ SAs, each visible by a UE with probability $P_b = 0.5$. \jcm{The employed XL-MIMO scenario is very similar to that assumed in \cite{Nishimura_2020}.} 


Simulations are generated in sequential RA blocks, which are represented by $10^4$ Monte Carlo realizations. In each round, new iUEs realize an access attempt following \eqref{eq: binomial distribution}. UEs that failed at the first attempt retransmit their RA pilots with probability $0.5$ in the next blocks. It is considered a failed access attempt if a UE do not succeed in a total of $10$ RA attempts, starting from its first transmission.

\subsection{Random Access Performance} 
\begin{figure}[htbp!]
\vspace{-2mm}
\centering
\includegraphics[trim={0mm 0mm 1mm 0mm},clip,width=.99\linewidth]{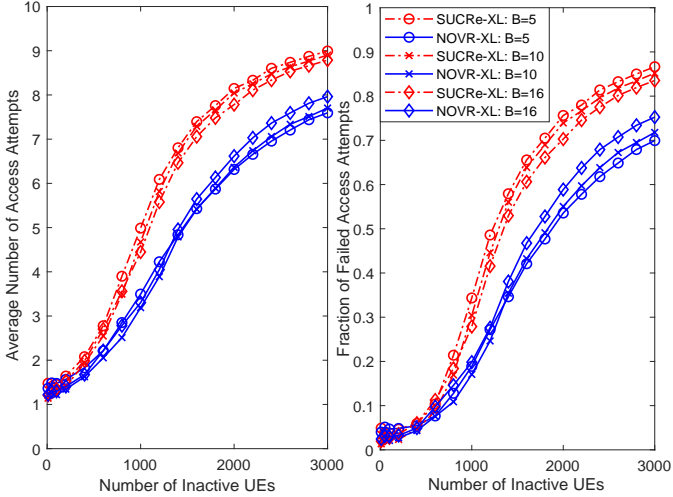}
\caption{Average number of access attempts and probability of failed access attempts as a function of the number of inactive UEs $K$, for $P_b = 0.5$, $P_a = 0.01$, $\tau_{\textsc{ra}}=10$ and $M=400$.}
\label{fig:RA_perf}
\end{figure}

We compare in Fig. \ref{fig:RA_perf} the average number of access attempts and the probability of failed access attempts as a function of the number of iUEs, considering our proposed NOVR-XL RA protocol, SUCRe-XL~\cite{Nishimura_2020}, and mSUCRe-XL as discussed in \jcm{Subsection \ref{sec:variant_RA_XL-MIMO}}. Note that SUCRe-XL and mSUCRe-XL achieve the same RA performance, differing only in terms of sum-rate. When obtaining these results we assumed that there is always an available PDP to be allocated for a UE that succeeded in the RA stage. Besides, for the proposed NOVR-XL RA protocol we assumed that the UL RA additional information vector of UE $k$ can be successfully decoded from SA $b$ if the UL SINR of $\widehat{\textbf{d}^{\rm ul}_{k}}$ in \eqref{eq:d_hat_gen}, as described in \eqref{eq:SINR_full}, is greater than a threshold of $0$~dB. In this case, the BS assumes that reliable channel estimates are available for this UE in this SA, evaluating $\widehat{\bf h}^{(b)}_k = \widehat{\bf y}^{(b)}_{r(k)}$ and $[\textbf{E}]_{b,k} = 1$. Furthermore, the UE $k$ succeeds in its RA attempt if the received DL message from step~2, $\widehat{\textbf{d}^{\rm dl}_{k'}}$ in \eqref{eq:step2:rx_signal_UE}, can be successfully decoded, as illustrated in {Fig.}~\ref{fig:flux}, assuming the same decoding threshold as in the uplink. We have chosen a threshold of $0$~dB for successfully decoding the messages since this is a reasonable value for decoding uplink control information (UCI) in the physical uplink control channel (PUCCH) for most allowed transmit formats in 5G \cite{3GPP_TS}. 

In \cite{Nishimura_2020}, authors assumed that the RA attempt is successful for SUCRe-XL if only one UE retransmits its chosen pilot in the step~3 of that protocol, or if the UEs retransmitting the same pilot have non-overlapping VRs. Herein, for a fair comparison, we have assumed that the UL signal received at the BS in step~3 of SUCRe-XL {protocol} is successfully decoded in SA $b$ if its SINR is above a threshold of $0$~dB. 
{As a result,} the channel estimates {processed in such} SA for this UE are used for precoding its DL signal in step~4. Finally, the UE has its RA attempt successful if the SINR of the DL signal received in step~4 is above a threshold of $0$~dB. In order to obtain the SINR of the UL signal received at the SA $b$ in step~3 of SUCRe-XL protocol, and the SINR of the DL signal received at the UE $k$ in step~4, we need to first define the set $\mathcal{R}_t$, composed by the indices of the UEs retransmitting the RA pilot $t$ in step~3 of that protocol. Besides, we define the set $\mathcal{R}$, composed by the indices of all UEs retransmitting any RA pilot in step~3, \emph{i.e.}, $\mathcal{R} = \mathcal{R}_1 \cup \mathcal{R}_2 \cup \ldots \mathcal{R}_{\tau_{\textsc{ra}}}$. Then, the SINR of the UL signal received at the SA $b$ in step~3 of SUCRe-XL protocol can be obtained by \eqref{eq:SINR_full}, but replacing $\mathcal{S}_t$ by $\mathcal{R}_t$, and $\mathcal{K}$ by $\mathcal{R}$. Finally, the subsets $\mathcal{C}$ and $\mathcal{I}_k$ can be obtained for this protocol in the same way as described for NOVR-XL, and \eqref{eq:SINR_DL_full} is readily applicable for obtaining the SINR of the DL signal received at the UE $k$ in step~4.

As one can see from Fig.~\ref{fig:RA_perf}, the proposed NOVR-XL protocol achieves significant improved RA performance in comparison with SUCRe-based {protocols}. For instance, {considering} $K = 2000$ iUEs, and $B=10$ SAs, the average number of access attempts decreases from {$\bar{\mu}_{\textsc{ra}}=8.064$  attempts} for SUCRe-XL to {$\bar{\mu}_{\textsc{ra}}= 6.363$ attempts} for NOVR-XL. In the same way, the probability of failed access attempts decreases from $0.7393$ to $0.5505$, representing a significant performance improvement of the NOVR-XL protocol in (over)crowded mMTC scenarios.

\vspace{2mm}
\noindent \textit{Remark 1}: It is noteworthy that the result discussed above is in accordance with the Markov's inequality \cite{Gubner}. Defining the random variable {$\mu_{\textsc{ra}}^{k}$} as the number of RA attempts the UE $k$ performed, the Markov bound can be readily applied as:
\begin{equation}\label{eq:Markov}
    {\rm Pr}(\mu_{\textsc{ra}}^k\geq 10) \leq \frac{\mathbb{E}[\mu_{\textsc{ra}}^k]}{10}.
\end{equation}
In \eqref{eq:Markov}, one can recognize ${\rm Pr}(\mu_{\textsc{ra}}^k\geq 10)$ as the probability of failed access attempt, and $\mathbb{E}[\mu_{\textsc{ra}}^k]$ as the average number of access attempts.

\subsection{Sum-Rate} 
In order to evaluate the sum-rate performance of the investigated protocols, we need first to obtain appropriate values for $\varphi_{\textsc{ra}}$, \emph{i.e.}, the number of channel uses spent per RA attempt of each RA protocol. In the SUCRe-XL protocol \cite{Nishimura_2020}, the four steps   present a very similar structure to that of medium access control in 5G, as described in \cite{3GPP_38321}. Therefore, we can use this document as reference. In SUCRe-XL, the UE transmits an RA pilot in step~1, spending $\tau_{\textsc{ra}}$ channel uses. The BS responds in step~2 with the precoded RA response, composed by the DL RA pilot, the RA resource network temporary identifier (RA-RNTI), and a timing advance (TA) information. While the DL RA pilot transmission spends $\tau_{\textsc{ra}}$ channel uses, the RA-RNTI and TA are of $16$ and $8$ bits length \cite{3GPP_38321}, respectively. For simplicity, we assume in our analysis that the modulation and coding scheme (MCS) employed is chosen resulting in a unitary MCS spectral efficiency\footnote{This means that the coding rate multiplied by the number of bits per modulated symbol is equal to $1$.}. Therefore, the number of channel uses in step~2 is $ {24+}\tau_{\textsc{ra}}$. The UE transmits in step~3 the repeated RA pilot ($\tau_{\textsc{ra}}$ channel uses) and the cell radio network temporary identifier (C-RNTI), of $16$ bits \cite{3GPP_38321}, such that the number of channel uses in this step is ${16+}\tau_{\textsc{ra}}$. Finally, in step~4, the BS responds with the C-RNTI and the contention resolution information (CRI), of 48 bits \cite{3GPP_38321}, such that the number of channel uses in this step is $64$. In summary, for SUCRe-XL, $\varphi^{\rm SUCRe-XL}_{\textsc{ra}} = 104 + 3 \tau_{\textsc{ra}}$.

For the mSUCRe-XL RA protocol discussed in Section~\ref{sec:variant_RA_XL-MIMO}, the signals transmitted in the four steps of the RA stage are very similar to that of SUCRe-XL, except that $\widehat{\bf v}_k$ is transmitted in step~3. Such estimate is a binary vector of length $B$, and thus we assume its transmission requires $B$ channel uses. Therefore, we have that $\varphi^{\rm mSUCRe-XL}_{\textsc{ra}} = 104 + B + 3 \tau_{\textsc{ra}}$.

Finally, for the proposed NOVR-XL protocol, in the RA stage we assume the UE transmits in step~1 its chosen RA pilot, an identifier information (ID) with $16$ bits, and $\widehat{\bf v}_k$, spending $16 + B + \tau_{\textsc{ra}}$ channel uses at this step. Then, the BS responds in step~2 with the identifier information, followed by TA and CRI, spending at this step $72$ channel uses. Therefore, we have that $\varphi^{\rm NOVR-XL}_{\textsc{ra}} = 88 + B + \tau_{\textsc{ra}}$. Table~\ref{tab:RAcu} summarizes the number of channel uses per RA attempt for each protocol under consideration. 

\begin{table}[!htbp]
\caption{Channel usage per attempt {for the three} RA protocols.}
\vspace{-2mm}
\centering
\small
\begin{tabular}{|c|c|c|c|}
\hline
\bf Step &\bf  SUCRe-XL & \bf mSUCRe-XL & \bf NOVR-XL\\
\hline
\hline
Step 1 & $\tau_{\textsc{ra}}$ (pilot) & $\tau_{\textsc{ra}}$ (pilot) & $\tau_{\textsc{ra}}$ (pilot) + \\
  &  &  & $B$ ($\widehat{{\bf v}}_k$) + \\
  &  &  & $16$ (ID) \\
\hline
Step 2 & $\tau_{\textsc{ra}}$ (pilot) + & $\tau_{\textsc{ra}}$ (pilot) + & $48$ (CRI) + \\
  & $16$ (RA-RNTI) + & $16$ (RA-RNTI) +  & $16$ (ID) + \\
  & $8$ (TA) & $8$ (TA) & $8$ (TA) \\
\hline
Step 3 & $\tau_{\textsc{ra}}$ (pilot) + & $\tau_{\textsc{ra}}$ (pilot) + &  \\
  & $16$ (C-RNTI) & $16$ (C-RNTI) + & $--$ \\
  &  & $B$ ($\widehat{{\bf v}}_k$) &  \\
\hline
Step 4 & $48$ (CRI) + & $48$ (CRI) + & $--$ \\
  & $16$ (C-RNTI) & $16$ (C-RNTI) &  \\
  \hline
\bf Total & $104 + 3 \tau_{\textsc{ra}}$ & $104 + B + 3 \tau_{\textsc{ra}}$ & $88 + B + \tau_{\textsc{ra}}$\\
\hline
\end{tabular}
\label{tab:RAcu}
\end{table}
\normalsize

\begin{figure}[htbp!]
\vspace{-2mm}
\centering
\includegraphics[trim={0mm 0mm 1mm 0mm},clip,width=1\linewidth]{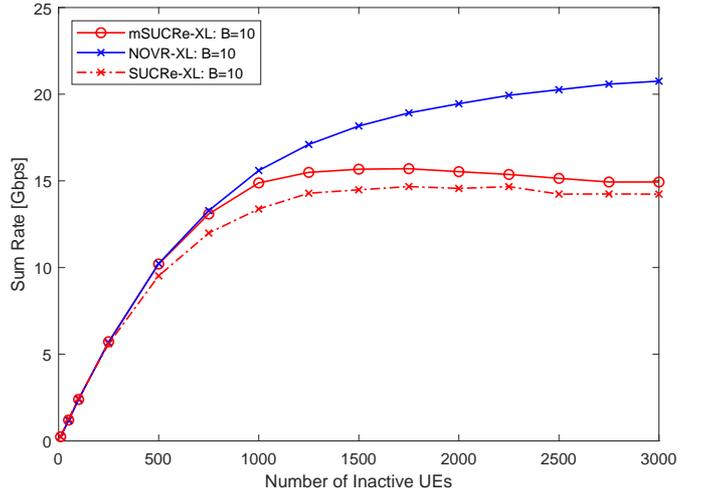}
\caption{Average sum-rate as a function of the number of inactive UEs $K$, for $B=10$ {SAs}, $P_b = 0.5$, $P_a = 0.01$, $\tau_{\textsc{ra}}=10$ and $M=400$ {antennas}.}
\label{fig:SR_perf}
\end{figure}

Given the $\varphi_{\textsc{ra}}$ values for the three investigated RA procedures, {Fig.}~\ref{fig:SR_perf} depicts the sum-rate performance obtained with each scheme, for $B=10$ SAs, $t_c = 1$ ms, ${\textsc{w}} = 20$ MHz, $T = 200$ channel uses, and {$\mu_{\textsc{pd}}^k = 10$} intervals, $\forall k \in \mathcal{A}$. As one can see, the proposed NOVR-XL protocol achieves the best sum-rate performance, with substantial improvements in comparison with mSUCRe-XL and SUCRe-XL. The Figure also shows that mSUCRe-XL outperforms SUCRe-XL, indicating that the application of our proposed scheduling procedure described in Algorithm \ref{alg:sched} 
{compensates the} overhead required by the transmission of $\widehat{\bf v}_k$. In terms of sum-rate, this occurs because more UEs can be scheduled in the same PDP sequence, requiring a smaller set of PDP sequences, and little training overhead since we assume $\tau_{\textsc{PD}} = |\mathcal{P}|$ in our system model, as described in Section \ref{sec:model}.

\begin{figure}[htbp!]
\vspace{-2mm}
\centering
\includegraphics[trim={1mm 1mm .5mm 0mm},clip,width=1.02\linewidth]{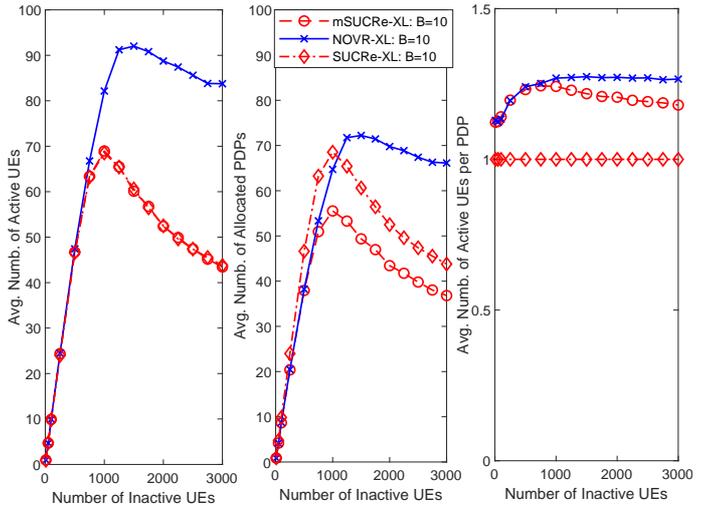}
\vspace{-5mm}
\caption{{Average number of \textbf{a)} active UEs, \textbf{b)} allocated PDPs, \textbf{c)} active UEs per allocated PDP, as a function of the number of inactive UEs $K$, {and assuming} $B=10$, $P_b = 0.5$, $P_a = 0.01$, $\tau_{\textsc{ra}}=10$, and $M=400$.}}
\label{fig:NAUE_perf}
\end{figure}

{Fig.} \ref{fig:NAUE_perf} analyses the effectiveness of each protocol in admitting UEs to the network, as well as efficiently scheduling PDP resources to them. The Figure depicts the average number of active UEs, the average number of allocated PDPs ($|\mathcal{P}| = \tau_{\textsc{PD}}$), and the average number of active UEs scheduled in the same PDP resource. One can see that the proposed NOVR-XL protocol is able to manage a substantially higher number of active UEs in the network, due to the better connectivity performance in the RA stage, as presented in {Fig.}~\ref{fig:RA_perf}. Furthermore, the Figure {reveals} that the {proposed} scheduling procedure for non-stationary XL-MIMO systems in Algorithm~\ref{alg:sched} is very effective in terms of simultaneously allocating the same PDP resource to different UEs, provided that they present non-overlapping VRs. In the case of NOVR-XL protocol, on average up to $1.274$ UEs can be scheduled in the same PDP resource. \jcm{The mSUCRe-XL RA protocol also achieves a good number of active UEs connected in the same PDP sequence for low/moderate number of iUEs, but its limited RA performance in (over)crowded scenarios reduces the number of active UEs in such cases.}

\section{Final Remarks}\label{sec:conclusion}
XL-MIMO systems are a new topology which promises appreciable performance improvements for {5G and 6G systems}. However, many difficulties have to be overcome in order to make this technology applicable in practice. This paper addresses the problems of RA and resource scheduling, proposing a \jcm{joint RA and scheduling} protocol which takes advantage of the random VRs of the UEs for improving both performance metrics. Our numerical results show that the connectivity performance is substantially improved, reducing the average number of access attempts and the probability of failed access attempts. Then, a detailed analysis of each RA protocol indicated the number of channel uses spent per RA attempt. This has allowed a sum-rate performance analysis, able to measure in a unified way the overhead of each protocol, its connectivity performance, and its capability of managing a large number of active UEs, requiring for this the \jcm{smallest} possible set of PDP sequences. Again, the proposed NOVR-XL protocol presented very improved performance results, showing that an increased  number of active UEs can be supported at the same time, {hence,} demanding  a smaller set of PDPs, taking advantage of the non-overlapping VRs of the UEs. \jcm{Therefore, our proposed protocol is able to reduce access latency in the same time it improves the XL-MIMO system sum-rate.}

 \vspace{-1mm}

\end{document}